\documentclass[12pt]{article}
\usepackage{graphicx}
\def\beq{\begin{equation}}
\def\eeq{\end{equation}}
\def\bea{\begin{eqnarray}}
\def\eea{\end{eqnarray}}
\def\nn{\nonumber}
\def\sss{\scriptscriptstyle}
\def\bd{B_d^0}
\def\bdbar{{\bar B}_d^0}
\def\bs{B_s^0}

\def\barp{{\raise.35ex\hbox
{${\sss (}$}}---{\raise.35ex\hbox{${\sss )}$}}}
\def\bdbarp{\hbox{$B_d$\kern-1.4em\raise1.4ex\hbox{\barp}}}
\def\bsbarp{\hbox{$B_s$\kern-1.4em\raise1.4ex\hbox{\barp}}}
\def\ks{K_{\sss S}}

\def\barpk{{\raise.35ex\hbox
{${\sss (}$}}--{\raise.35ex\hbox{${\sss )}$}}}
\def\kbarp{\hbox{$K$\kern-0.9em\raise1.4ex\hbox{\barpk}}}
\def\bbarp{\hbox{$B$\kern-0.9em\raise1.4ex\hbox{\barpk}}}
\def\roughly#1{\mathrel{\raise.3ex\hbox
{$#1$\kern-.75em\lower1ex\hbox{$\sim$}}}}

\def\bvv{B \to V_1 V_2}
\def\epjc#1#2#3{{\it Eur.\ Phys.\ J.}\ {\bf C#1}, #3, #2}

\def\plb#1#2#3{{\it Phys.\ Lett.} {\bf #1B}, #3, #2}
\def\prd#1#2#3{{\it Phys.\ Rev.} {\bf D#1}, #3, #2}
\def\newprd#1#2#3{{\it Phys.\ Rev.} {\bf D#1}: #3, #2}
\def\prl#1#2#3{{\it Phys.\ Rev.\ Lett.} {\bf #1}, #3, #2}
\def\newprl#1#2#3{{\it Phys.\ Rev.\ Lett.} {\bf #1}: #3, #2}

\newread\epsffilein 
\newif\ifepsffileok 
\newif\ifepsfbbfound 
\newif\ifepsfverbose 
\newdimen\epsfxsize 
\newdimen\epsfysize 
\newdimen\epsftsize 
\newdimen\epsfrsize 
\newdimen\epsftmp 
\newdimen\pspoints 
\def\epsfbox#1{\global\def\epsfllx{72}\global\def\epsflly{72}%
 \global\def\epsfurx{540}\global\def\epsfury{720}%
 \def\lbracket{[}\def\testit{#1}\ifx\testit\lbracket
 \let\next=\epsfgetlitbb\else\let\next=\epsfnormal\fi\next{#1}}%
\def\epsfgetlitbb#1#2 #3 #4 #5]#6{\epsfgrab #2 #3 #4 #5 .\\%
 \epsfsetgraph{#6}}%
\def\epsfnormal#1{\epsfgetbb{#1}\epsfsetgraph{#1}}%
\def\epsfgetbb#1{%
\openin\epsffilein=#1
\ifeof\epsffilein\errmessage{I couldn't open #1, will ignore it}\else
 {\epsffileoktrue \chardef\other=12
 \def\do##1{\catcode`##1=\other}\dospecials \catcode`\ =10
 \loop
 \read\epsffilein to \epsffileline
 \ifeof\epsffilein\epsffileokfalse\else
 \expandafter\epsfaux\epsffileline:. \\%
 \fi
 \ifepsffileok\repeat
 \ifepsfbbfound\else
 \ifepsfverbose\message{No bounding box comment in #1; using defaults}\fi\fi
 }\closein\epsffilein\fi}%
\def\epsfclipstring{}
\def\epsfsetgraph#1{%
 \epsfrsize=\epsfury\pspoints
 \advance\epsfrsize by-\epsflly\pspoints
 \epsftsize=\epsfurx\pspoints
 \advance\epsftsize by-\epsfllx\pspoints
 \epsfxsize\epsfsize\epsftsize\epsfrsize
 \ifnum\epsfxsize=0 \ifnum\epsfysize=0
 \epsfxsize=\epsftsize \epsfysize=\epsfrsize
 \epsfrsize=0pt
 \else\epsftmp=\epsftsize \divide\epsftmp\epsfrsize
 \epsfxsize=\epsfysize \multiply\epsfxsize\epsftmp
 \multiply\epsftmp\epsfrsize \advance\epsftsize-\epsftmp
 \epsftmp=\epsfysize
 \loop \advance\epsftsize\epsftsize \divide\epsftmp 2
 \ifnum\epsftmp>0
 \ifnum\epsftsize<\epsfrsize\else
 \advance\epsftsize-\epsfrsize \advance\epsfxsize\epsftmp \fi
 \repeat
 \epsfrsize=0pt
 \fi
 \else \ifnum\epsfysize=0
 \epsftmp=\epsfrsize \divide\epsftmp\epsftsize
 \epsfysize=\epsfxsize \multiply\epsfysize\epsftmp
 \multiply\epsftmp\epsftsize \advance\epsfrsize-\epsftmp
 \epsftmp=\epsfxsize
 \loop \advance\epsfrsize\epsfrsize \divide\epsftmp 2
 \ifnum\epsftmp>0
 \ifnum\epsfrsize<\epsftsize\else
 \advance\epsfrsize-\epsftsize \advance\epsfysize\epsftmp \fi
 \repeat
 \epsfrsize=0pt
 \else
 \epsfrsize=\epsfysize
 \fi
 \fi
 \ifepsfverbose\message{#1: width=\the\epsfxsize, height=\the\epsfysize}\fi
 \epsftmp=10\epsfxsize \divide\epsftmp\pspoints
 \vbox to\epsfysize{\vfil\hbox to\epsfxsize{%
 \ifnum\epsfrsize=0\relax
 \includegraphics{#1}%
 \else
 \epsfrsize=10\epsfysize \divide\epsfrsize\pspoints
 \includegraphics{#1}%
 \fi
 \hfil}}%
\global\epsfxsize=0pt\global\epsfysize=0pt}%
 {\catcode`\%=12 \global\let\epsfpercent=
\long\def\epsfaux#1#2:#3\\{\ifx#1\epsfpercent
 \def\testit{#2}\ifx\testit\epsfbblit
 \epsfgrab #3 . . . \\%
 \epsffileokfalse
 \global\epsfbbfoundtrue
 \fi\else\ifx#1\par\else\epsffileokfalse\fi\fi}%
\def\epsfempty{}%
\def\epsfgrab #1 #2 #3 #4 #5\\{%
\global\def\epsfllx{#1}\ifx\epsfllx\epsfempty
 \epsfgrab #2 #3 #4 #5 .\\\else
 \global\def\epsflly{#2}%
 \global\def\epsfurx{#3}\global\def\epsfury{#4}\fi}%
\def\epsfsize#1#2{\epsfxsize}
\pagestyle{plain}

\begin{document}

\begin{flushright}  
UdeM-GPP-TH-04-117 \\
McGill/04/02 \\
IMSc-2004/02/01 \\
\end{flushright}

\begin{center}
\bigskip
{\Large \bf \boldmath Bounds on New Physics from $\bvv$ Decays}
\end{center}

\begin{center}
{\large David London $^{a,b,}$\footnote{london@lps.umontreal.ca}, Nita
Sinha $^{c,}$\footnote{nita@imsc.res.in} and Rahul Sinha
$^{c,}$\footnote{sinha@imsc.res.in}}
\end{center}

\begin{flushleft}
~~~~~~~~~~~$a$: {\it Physics Department, McGill University,}\\
~~~~~~~~~~~~~~~{\it 3600 University St., Montr\'eal QC, Canada H3A 2T8}\\
~~~~~~~~~~~$b$: {\it Laboratoire Ren\'e J.-A. L\'evesque, 
Universit\'e de Montr\'eal,}\\
~~~~~~~~~~~~~~~{\it C.P. 6128, succ. centre-ville, Montr\'eal, QC,
Canada H3C 3J7} \\
~~~~~~~~~~~$c$: {\it Institute of Mathematical Sciences, C. I. T
 Campus,}\\
~~~~~~~~~~~~~~~{\it Taramani, Chennai 600 113, India}
\end{flushleft}

\begin{center} 
\bigskip (\today)
\vskip0.5cm
{\Large Abstract\\}
\vskip3truemm
\parbox[t]{\textwidth} {We consider the possibility that physics
beyond the standard model contributes to the decays $\bvv$, where
$V_1$ and $V_2$ are vector mesons. We show that a time-dependent
angular analysis of $\bvv$ decays provides many tests for this new
physics (NP). Furthermore, although one cannot solve for the NP
parameters, we show that this angular analysis allows one to put
bounds on these parameters. This can be useful in estimating the scale
of NP, and can tell us whether any NP found directly at future
high-energy colliders can be responsible for effects seen in $\bvv$
decays.}
\end{center}

\thispagestyle{empty}
\newpage
\setcounter{page}{1}
\baselineskip=14pt

\section{Introduction}

Within the standard model (SM), a complex phase in the
Cabibbo-Kobayashi-Maskawa (CKM) quark mixing matrix is responsible for
CP violation \cite{pdg}. By studying CP-violating processes in the $B$
system, one can test this explanation. If any discrepancy with the SM
predictions is found, this would be evidence for physics beyond the
SM.

There are a great many tests for the presence of new physics (NP) in
$B$ decays \cite{CPreview}. Should a signal for NP be found, there are
basically two ways to proceed. One can examine various models of
physics beyond the SM to see whether a particular model can account
for the experimental results. Alternatively, one can perform a
model-independent analysis to learn about general properties of the NP
responsible for the signal. Most theoretical work has concentrated on
the first approach.

For example, within the SM, the CP-violating asymmetries in $\bd(t)
\to J/\psi \ks$ and $\bd(t) \to \phi \ks$ both measure the CP phase
$\beta$, to a good approximation \cite{phiKs}. However, although the
BaBar measurement of the CP asymmetry in $\bd(t) \to \phi \ks$ agrees
with that found in $\bd(t) \to J/\psi \ks$ (within errors), the Belle
measurement disagrees at the level of $3.5\sigma$ \cite{browder}.
This suggests that physics beyond the SM --- specifically new decay
amplitudes in $B \to \phi K$ --- may be present. In light of this,
many papers have been written to show how particular models of NP can
account for this discrepancy \cite{phiKsNP}. On the other hand, only
two papers contain a model-independent analysis of $\bd(t) \to \phi
\ks$ \cite{phiKmodind} (and even here some theoretical numerical input
is required).

In this paper, we show how model-independent information about new
physics can be obtained from an angular analysis of $\bvv$ decays,
where $V_1$ and $V_2$ are vector mesons. This method is applicable to
those $\bvv$ decays in which (i) ${\overline{V}}_1 {\overline{V}}_2 =
V_1 V_2$, so that this final state is accessible to both $B^0$ and
${\bar B}^0$, and (ii) a single decay amplitude dominates in the SM.
The only theoretical assumption we make is that there is only a single
NP amplitude, with a different weak phase from that of the SM
amplitude, contributing to these decays. In the event that a signal
for NP is found, we demonstrate that one can place {\it lower} bounds
on the NP parameters \cite{LSSbounds}.

If physics beyond the SM contributes to $\bd(t) \to \phi \ks$, there
should also be NP signals in the corresponding $\bvv$ decay, $\bd(t)
\to \phi K^{*0}$. Our method can be used in this situation to get
information about the NP. It can also be applied to $\bd(t) \to J/\psi
K^{*0}$, $\bd(t) \to K^{*0} {\bar K}^{*0}$, $\bs(t) \to J/\psi \phi$,
etc., should NP signals be found in these decays\footnote{Our analysis
treats only the situation where there are additional NP decay
amplitudes; it does not apply to the case where the NP appears only in
$B^0$--${\bar B}^0$ mixing.}.

Any new-physics effects in $B$ decays are necessarily virtual. On the
other hand, future experiments at the Large Hadron Collider (LHC) and
at a linear $e^+ e^-$ collider (GLC) will make direct searches for
such NP. Should NP be found in both $\bvv$ decays and at the LHC/GLC,
the bounds from the angular analysis can tell us whether the NP seen
at LHC/GLC can be responsible for the effects in $\bvv$ decays.

We begin in Sec.~2 by describing the theoretical framework of our
method. Signals of new physics are examined in Sec.~3. The main
results --- how to place bounds on the theoretical NP parameters ---
are presented in Sec.~4. We discuss and summarize these results in
Sec.~5.

\section{Theoretical Framework}

Consider a $\bvv$ decay which is dominated by a single weak decay
amplitude within the SM. This holds for processes which are described
by the quark-level decays ${\bar b} \to {\bar c} c {\bar s}$, ${\bar
b} \to {\bar s} s {\bar s}$ or ${\bar b} \to {\bar s} d {\bar d}$. In
all cases, the weak phase of the SM amplitude is zero in the standard
parametrization \cite{pdg}. Suppose now that there is a single
new-physics amplitude, with a different weak phase, that contributes
to the decay. The decay amplitude for each of the three possible
helicity states may be written as
\bea
A_\lambda \equiv Amp (\bvv)_\lambda &=& a_\lambda e^{i
\delta_\lambda^a} + b_\lambda e^{i\phi} e^{i \delta_\lambda^b} ~,
\nn\\
{\bar A}_\lambda \equiv Amp ({\bar B} \to (V_1 V_2)_\lambda &=&
a_\lambda e^{i \delta_\lambda^a} + b_\lambda e^{-i\phi} e^{i
\delta_\lambda^b} ~,
\label{amps}
\eea
where $a_\lambda$ and $b_\lambda$ represent the SM and NP amplitudes,
respectively, $\phi$ is the new-physics weak phase, the
$\delta_\lambda^{a,b}$ are the strong phases, and the helicity index
$\lambda$ takes the values $\left\{ 0,\|,\perp \right\}$. Using CPT
invariance, the full decay amplitudes can be written as
\bea
{\cal A} &=& Amp (\bvv) = A_0 g_0 + A_\| g_\| + i \, A_\perp
g_\perp~, \nn\\
{\bar{\cal A}} &=& Amp ({\bar B} \to V_1 V_2) = {\bar A}_0 g_0 + {\bar
A}_\| g_\| - i \, {\bar A}_\perp g_\perp~,
\label{fullamps}
\eea
where the $g_\lambda$ are the coefficients of the helicity amplitudes
written in the linear polarization basis. The $g_\lambda$ depend only
on the angles describing the kinematics \cite{glambda}. 

Note that it is not a strong assumption to consider a single NP
amplitude. First, the new physics is expected to be heavy, so that all
strong phases $\delta_\lambda$ should be small. In this case, since
the $\delta_\lambda$ are all of similar size, our parametrization
above is adequate. Second, if it happens that this is not the case,
and there are several different contributing NP amplitudes, our
analysis pertains to the dominant signal. Finally, if all the NP
amplitudes are of the same size, our approach provides an
order-of-magnitude estimate for the size of new physics.

Eqs.~(\ref{amps}) and (\ref{fullamps}) above enable us to write the
time-dependent decay rates as
\beq
\Gamma(\bbarp(t) \to V_1V_2) = e^{-\Gamma t} \sum_{\lambda\leq\sigma}
\Bigl(\Lambda_{\lambda\sigma} \pm \Sigma_{\lambda\sigma}\cos(\Delta M
t) \mp \rho_{\lambda\sigma}\sin(\Delta M t)\Bigr) g_\lambda g_\sigma
~.
\label{decayrates}
\eeq
Thus, by performing a time-dependent angular analysis of the decay
$B(t) \to V_1V_2$, one can measure 18 observables. These are:
\bea
\Lambda_{\lambda\lambda}=\displaystyle
\frac{1}{2}(|A_\lambda|^2+|{\bar A}_\lambda|^2),~~&&
\Sigma_{\lambda\lambda}=\displaystyle
\frac{1}{2}(|A_\lambda|^2-|{\bar A}_\lambda|^2),\nn \\[1.ex]
\Lambda_{\perp i}= -\!{\rm Im}({ A}_\perp { A}_i^* \!-\! {\bar
A}_\perp {{\bar A}_i}^* ),
&&\Lambda_{\| 0}= {\rm Re}(A_\| A_0^*\! +\! {\bar A}_\| {{\bar A}_0}^*
), \nn \\[1.ex]
\Sigma_{\perp i}= -\!{\rm Im}(A_\perp A_i^*\! +\! {\bar A}_\perp
{{\bar A}_i}^* ),
&&\Sigma_{\| 0}= {\rm Re}(A_\| A_0^*\!-\! {\bar A}_\| {{\bar A}_0}^*
),\nn\\[1.ex]
\rho_{\perp i}\!=\! {\rm Re}\!\Bigl(\frac{q}{p} \!\bigl[A_\perp^*
{\bar A}_i\! +\! A_i^* {\bar A}_\perp\bigr]\Bigr),
&&\rho_{\perp \perp}\!=\! {\rm Im}\Bigl(\frac{q}{p}\, A_\perp^*
{\bar A}_\perp\Bigr),\nn\\[1.ex]
\rho_{\| 0}\!=\! -{\rm Im}\!\Bigl(\frac{q}{p}[A_\|^* {\bar A}_0\! +
\!A_0^* {\bar A}_\| ]\Bigr),
&&\rho_{ii}\!=\! -{\rm Im}\!\Bigl(\frac{q}{p} A_i^* {\bar A}_i\Bigr),
  \label{eq:obs}
\eea
where $i=\{0,\|\}$. In the above, $q/p$ is the weak phase factor
associated with $B$--${\bar B}$ mixing. For $\bd$ mesons, $q/p =
\exp({-2\,i\beta})$, while $q/p = 1$ for $\bs$ mesons. Henceforth we
concentrate on the decays of $\bd$ mesons, though our results can
easily be adapted to $\bs$ decays. Note that $\beta$ may include NP
effects in $\bd$--$\bdbar$ mixing. Note also that the signs of the
various $\rho_{\lambda\lambda}$ terms depend on the CP-parity of the
various helicity states. We have chosen the sign of $\rho_{00}$ and
$\rho_{\|\|}$ to be $-1$, which corresponds to the final state $\phi
K^*$.

Not all of the 18 observables are independent. There are a total of
six amplitudes describing $\bvv$ and ${\bar B} \to V_1 V_2$ decays
[Eq.~(\ref{amps})]. Thus, at best one can measure the magnitudes and
relative phases of these six amplitudes, giving 11 independent
measurements.

The 18 observables given above can be written in terms of 13
theoretical parameters: three $a_\lambda$'s, three $b_\lambda$'s,
$\beta$, $\phi$, and five strong phase differences defined by
$\delta_\lambda \equiv \delta_\lambda^b - \delta_\lambda^a$, $\Delta_i
\equiv \delta_\perp^a - \delta_i^a$. The explicit expressions for the
observables are as follows:
\begin{eqnarray}
\Lambda_{\lambda\lambda} & = & a_\lambda^2 + b_\lambda^2 + 2 a_\lambda
b_\lambda \cos\delta_\lambda \cos\phi ~, \nn\\
\Sigma_{\lambda\lambda}  & = & - 2 a_\lambda b_\lambda
\sin\delta_\lambda \sin\phi ~, \nn\\
\Lambda_{\perp i} & = & 2 \left[ a_\perp b_i \cos(\Delta_i - \delta_i)
- a_i b_\perp \cos(\Delta_i + \delta_\perp) \right] \sin\phi ~,\nn\\
\Lambda_{\| 0} & = & 2 \left[ a_\| a_0 \cos(\Delta_0 - \Delta_\|) +
a_\| b_0 \cos(\Delta_0 - \Delta_\| - \delta_0) \cos\phi \right. \nn\\
& & ~~ \left. + ~ a_0 b_\| \cos(\Delta_0 - \Delta_\| + \delta_\|)
\cos\phi + b_\| b_0 \cos(\Delta_0 - \Delta_\| + \delta_\| - \delta_0)
\right] ~,\nn\\
\Sigma_{\perp i} & = & -2 \left[ a_\perp a_i \sin \Delta_i + a_\perp
b_i \sin(\Delta_i - \delta_i) \cos\phi \right. \nn\\
& & ~~ \left. + ~ a_i b_\perp \sin(\Delta_i + \delta_\perp) \cos\phi +
b_\perp b_i \sin (\Delta_i + \delta_\perp - \delta_i) \right] ~, \nn\\
\Sigma_{\| 0} & = & 2 \left[ a_\| b_0 \sin(\Delta_0 - \Delta_\| -
\delta_0) - a_0 b_\| \sin(\Delta_0 - \Delta_\| + \delta_\|) \right]
\sin\phi ~, \nn\\
\rho_{ii} & = & a_i^2 \sin 2\beta + 2 a_i b_i \cos\delta_i \sin(2
\beta + \phi) + b_i^2 \sin(2\beta + 2 \phi) ~, \nn\\
\rho_{\perp\perp} & = & - a_\perp^2 \sin 2\beta - 2 a_\perp b_\perp
\cos\delta_\perp \sin(2 \beta + \phi) - b_\perp^2 \sin(2\beta + 2
\phi) ~, \nn\\
\rho_{\perp i} & = & 2 \left[ a_i a_\perp \cos \Delta_i \cos 2\beta +
a_\perp b_i \cos(\Delta_i - \delta_i) \cos(2 \beta + \phi)
\right. \nn\\
& & ~~ + ~ a_i b_\perp \cos(\Delta_i + \delta_\perp) \cos(2 \beta
+ \phi) \nn\\
& & ~~ \left. + ~ b_i b_\perp \cos(\Delta_i + \delta_\perp - \delta_i)
\cos(2\beta + 2\phi) \right] ~,\nn\\
\rho_{\| 0} & = & 2 \left[ a_0 a_\| \cos(\Delta_0 - \Delta_\|) \sin
2\beta + a_\| b_0 \cos(\Delta_0 -\Delta_\| - \delta_0) \sin(2 \beta +
\phi) \right. \nn\\
& & ~~ + ~ a_0 b_\| \cos(\Delta_0 - \Delta_\| + \delta_\|) \sin(2
\beta + \phi) \nn\\
& & ~~ \left. + ~ b_0 b_\| \cos(\Delta_0 - \Delta_\| + \delta_\| -
\delta_0) \sin(2\beta + 2\phi) \right] ~.
\label{observables}
\end{eqnarray}
In subsequent sections, we will work extensively with these
expressions.

It is straightforward to see that, in the presence of new physics, one
cannot extract the phase $\beta$. There are 11 independent
observables, but 13 theoretical parameters. Since the number of
measurements is fewer than the number of parameters, one cannot
express any of the theoretical unknowns purely in terms of
observables. In particular, it is impossible to extract $\beta$
cleanly. Nevertheless, we will show that the angular analysis does
allow one to obtain significant {\it lower bounds} on the NP
parameters, as well as on the deviation of $\beta$ from its measured
value.

In our analysis, we usually assume that $\beta$ has not been measured
independently, so that there are indeed 13 unknown theoretical
parameters. However, this might not be the case. For example, the
decay $\bd(t) \to J/\psi \ks$ (or $\bd(t) \to J/\psi K^{*0}$) is
dominated by the tree contribution. Even if there is new physics in
the ${\bar b} \to {\bar c} c {\bar s}$ penguin amplitude, its effect
will probably be very small. If it is found experimentally that this
is so (e.g.\ using the NP signals discussed in the next section), the
measurement of the CP asymmetry in this mode gives the true (SM) value
of $\beta$. This can then be used as an input for other modes, such as
$\bd(t) \to \phi K^{*0}$. In this case there are only 12 theoretical
parameters, and the analysis simplifies. We will comment on this
possibility in Sec.~4.6.

\section{Signals of New Physics}

As mentioned in the introduction, lower bounds on new-physics
parameters are possible only if there is a signal of physics beyond
the SM. In this section, we discuss the possible new-physics signals
in $\bvv$ decays.

In the absence of NP, the $b_\lambda$ are zero in Eq.~(\ref{amps}).
The number of parameters is then reduced from 13 to 6: three
$a_\lambda$'s, two strong phase differences ($\Delta_i$), and
$\beta$. It is straightforward to show that all six parameters can be
determined cleanly in terms of observables [Eq.~(\ref{observables})].
However, there are a total of 18 observables. Thus, there must exist
12 relations among the observables in the absence of NP. These are:
\bea
&& \Sigma_{\lambda\lambda}= \Lambda_{\perp i}= \Sigma_{\| 0}=0 ~,
\nn\\
&& \frac{\rho_{ii}}{\Lambda_{ii}} =
-\frac{\rho_{\perp\perp}}{\Lambda_{\perp\perp}} =
\frac{\rho_{\|0}}{\Lambda_{\| 0}} ~, \nn \\
&& \Lambda_{\|0}=\frac{1}{2\Lambda_{\perp\perp}}\Bigl[
\frac{\Lambda_{\lambda\lambda}^2\rho_{\perp 0}
\rho_{\perp\|}+\Sigma_{\perp 0}\Sigma_{\perp
\|}(\Lambda_{\lambda\lambda}^2 -\rho_{\lambda\lambda}^2)}
{\Lambda_{\lambda\lambda}^2-\rho_{\lambda\lambda}^2}\Bigr] ~, \nn \\
&& \frac{\rho_{\perp
i}^2}{4\Lambda_{\perp\perp}\Lambda_{ii}-\Sigma_{\perp
i}^2}=\frac{\Lambda_{\perp\perp}^2
-\rho_{\perp\perp}^2}{\Lambda_{\perp\perp}^2} ~.
\label{eq:no_np}
\eea
The key point is the following: {\em the violation of any of the above
relations will be a smoking-gun signal of NP}. We therefore see that
the angular analysis of $\bvv$ decays provides numerous tests for the
presence of NP\footnote{Note that, despite the many tests, it is still
possible for the NP to remain hidden. If the three strong phase
differences $\delta_\lambda$ vanish, and the ratio $r_\lambda \equiv
b_\lambda/a_\lambda$ is the same for all helicities, i.e. $r_0 = r_\|
= r_\perp$, then it is easy to show that the relations in
Eq.~(\ref{eq:no_np}) are all satisfied. Thus, if these very special
conditions happen to hold, the angular analysis of $\bvv$ would show
no signal for NP even if it is present, and the measured value of
$\beta$ would not correspond to its true (SM) value. Still, we should
stress that it is highly unlikely that the NP parameters should
respect such a singular situation.}.

Since there are 11 independent observables and 6 parameters in the SM,
one might expect that only 5 tests are needed to verify the presence
of NP. However, since the equations in Eq.~(\ref{observables}) are
nonlinear, this logic can fail: if the SM parameters take certain
special values, more tests are needed. For example, suppose that $b_\|
= b_\perp = 0$ and $\delta_0 = 0$. Since $b_0 \ne 0$, NP is
present. We have $\Sigma_{\lambda\lambda} = \Lambda_{\perp \|} =
0$. If $\Delta_0$ takes the value $\pi/2$, we will also find that
$\Lambda_{\perp 0} = 0$. Thus, despite the presence of NP, 5 of the 12
tests above agree with the SM. In this case, further tests are needed
to confirm the fact that NP is present. In the most general case, {\it
all} 12 tests above are needed to search for NP. (In any event,
because it is not known a-priori which observables will be measured,
it is important to have a list of all NP tests.)

We should stress here that the list of NP signals is independent of
the parametrization of new physics. That is, even if there are several
contributing amplitudes, the NP can still be discovered through the
tests in Eq.~(\ref{eq:no_np}). Furthermore, even in this general case,
it is necessary to perform all 12 tests in order to show that NP is
not present.

The observable $\Lambda_{\perp i}$ deserves special attention. It is
the coefficient of the T-odd ``triple product'' in $\bvv$ decays,
${\vec q} \cdot ({\vec\varepsilon}_1 \times {\vec\varepsilon}_2)$,
where ${\vec q}$ is the momentum of one of the final vector mesons in
the rest frame of the $B$, and ${\vec\varepsilon}_{1,2}$ are the
polarizations of $V_1$ and $V_2$ \cite{DattaLondon}. From
Eq.~(\ref{observables}), one sees that even if the strong phase
differences vanish, $\Lambda_{\perp i}$ is nonzero in the presence of
new physics ($\phi\ne 0$), in contrast to the direct CP asymmetries
(proportional to $\Sigma_{\lambda\lambda}$). This is due to the fact
that the $\perp$ helicity is CP-odd, while the $0$ and $\|$ helicities
are CP-even. Thus, $\perp$--$0$ and $\perp$--$\|$ interferences
include an additional factor of `$i$' in the full decay amplitudes
[Eq.~(\ref{fullamps})], which leads to the cosine dependence on the
strong phases.

Although the reconstruction of the full $\bd(t)$ and $\bdbar(t)$ decay
rates in Eq.~(\ref{decayrates}) requires both tagging and
time-dependent measurements, the $\Lambda_{\lambda\sigma}$ terms
remain even if the two rates for $\bd(t)$ and $\bdbar(t)$ decays are
added together. Note also that these terms are time-independent.
Therefore, {\it no tagging or time-dependent measurements are needed
to extract $\Lambda_{\perp i}$}. It is only necessary to perform an
angular analysis of the final state $V_1 V_2$. Thus, this measurement
can even be made at a symmetric $B$-factory.

The decays of charged $B$ mesons to vector-vector final states are
even simpler to analyze since no mixing is involved. One can in
principle combine charged and neutral $B$ decays to increase the
sensitivity to new physics. For example, for $B \to J/\psi K^*$
decays, one simply performs an angular analysis on all decays in which
a $J/\psi$ is produced accompanied by a charged or neutral $K^*$. A
nonzero value of $\Lambda_{\perp i}$ would be a clear signal for new
physics \cite{FPCP}.

\section{Bounds on the Theoretical Parameters}

In this section we explore the constraints on the size of new physics,
assuming that a NP signal is observed in $\bvv$. As we have shown, the
amplitudes are written in terms of 13 theoretical parameters
(including $\beta$), but there are only 11 independent observables.
Since the number of unknowns is greater than the number of
observables, naively one would think that it is not possible to obtain
any information about the NP parameters. However, since the
expressions for the observables in terms of the theoretical parameters
are nonlinear [Eq.~(\ref{observables})], it is in fact possible to
obtain {\it bounds} on the NP parameters. One can even put a lower
bound on the difference between the measured value of $\beta$ (which
is affected by the presence of NP) and its true (SM) value.

The first step is to reduce the number of unknowns in the expressions
for the observables. That is, even though one cannot solve for the
theoretical parameters in terms of observables, one can obtain a
partial solution, in which observables are written in terms of a
smaller number of parameters plus other observables.

For $\bvv$ decays, the analogue of the usual direct CP asymmetry
$a^{CP}_{dir}$ is $a_{\lambda}^{dir} \equiv
\Sigma_{\lambda\lambda}/\Lambda_{\lambda\lambda}$, which is
helicity-dependent. We define the related quantity,
\beq
y_\lambda \equiv \sqrt{1 -
\Sigma_{\lambda\lambda}^2/\Lambda_{\lambda\lambda}^2}~.
\eeq
The measured value of $\sin 2\beta$ can also depend on the helicity of
the final state: $\rho_{\lambda\lambda}$ can be recast in terms of a
measured weak phase $2\beta^{meas}_{\lambda}$, defined as
\bea
\sin\, 2\,\beta^{meas}_{\lambda} \equiv \frac{\pm
\rho_{\lambda\lambda}}{\sqrt{\Lambda^2_{\lambda\lambda}-
\Sigma^2_{\lambda\lambda}}} ~,
\label{betameasdef}
\eea
where the $+$ $(-)$ sign corresponds to $\lambda=0,\|$ ($ \perp$).

It is possible to express the 9 theoretical parameters $a_\lambda$,
$b_\lambda$ and $\delta_\lambda$ in terms of the 9 observables
$\Lambda_{\lambda\lambda}$, $\Sigma_{\lambda\lambda}$, and
$\rho_{\lambda\lambda}$, and the parameters $\beta$ and $\phi$. The
other observables can in turn be expressed in terms of
$\Lambda_{\lambda\lambda}$, $\Sigma_{\lambda\lambda}$, and
$\rho_{\lambda\lambda}$, along with the three theoretical parameters
$\beta + \phi$ and $\Delta_i$. Using the expressions for
$\Lambda_{\lambda\lambda}$, $\Sigma_{\lambda\lambda}$ and
$\beta^{meas}_\lambda$ above, one can express $a_\lambda$ and
$b_\lambda$ as follows:
\bea
\label{eq:a-b}
2\,a_\lambda^2\,\sin^2\phi &=& \Lambda_{\lambda\lambda}
\Big(1-y_\lambda\cos(2\beta^{meas}_\lambda-2\beta-2\phi)\Big) ~, \\
\label{eq:a-b-2}
2\,b_\lambda^2\,\sin^2\phi &=& \Lambda_{\lambda\lambda}
\Big(1-y_\lambda\cos(2\beta^{meas}_\lambda-2\beta)\Big) ~.
\end{eqnarray}
These expression will play a critical role in the derivation of bounds
on the NP parameters. 

The seemingly impossible task of eliminating 10 combinations of the
theoretical parameters $a_\lambda$, $b_\lambda$, $\delta_\lambda$,
$\beta$ and $\phi$ in terms of the observables
$\Lambda_{\lambda\lambda}$, $\Sigma_{\lambda\lambda}$ and
$\rho_{\lambda\lambda}$, and variable $\beta + \phi$ becomes possible
by using the following relation:
\bea
  \frac{b_\lambda}{a_\lambda} \cos\delta_\lambda\,\cos\phi 
\hskip-2.7truemm &=& \hskip-2.7truemm 
\frac{-2\Lambda_{\lambda\lambda} \cos^2\phi
      +y_\lambda\,\Lambda_{\lambda\lambda}\,
     (\cos(2\beta^{meas}_\lambda-2\beta-2\phi)
      +\cos(2\beta^{meas}_\lambda-2\beta))}
     {2 \Lambda_{\lambda\lambda}(1-y_\lambda
                \cos(2\beta^{meas}_\lambda-2\beta-2\phi))}\nn \\
\hskip-2.7truemm &=& \hskip-2.7truemm 
-\cos^2\phi\Bigg(1+\frac{y_\lambda \sin(2\beta^{meas}_\lambda
      -2\beta-2\phi)\tan\phi}{1-y_\lambda\cos(2\beta^{meas}_\lambda
      -2\beta-2\phi)}\Bigg) ~,
\label{eq:Lcosphi-1}
\eea
where we have used the expression for $\Lambda_{\lambda\lambda}$ given
in Eq.~(\ref{observables}). We introduce a compact notation to express
Eq.~(\ref{eq:Lcosphi-1}) by defining
\bea
P_\lambda^2 &\equiv& \Lambda_{\lambda\lambda}(1-y_\lambda\,
       \cos(2\beta^{meas}_\lambda-2\beta-2\phi))~,\\
\xi_\lambda &\equiv& \frac{\Lambda_{\lambda\lambda}\, y_\lambda\,
       \sin(2\beta^{meas}_\lambda-2\beta-2\phi)}{P_\lambda^2}~.
\label{eq:defs}
\eea
This results in
\beq
\frac{b_\lambda}{a_\lambda} \cos\delta_\lambda\,\cos\phi =
-\cos^2\phi-\cos\phi\sin\phi\,\xi_\lambda
\label{eq:Lcosphi}
\eeq
Similarly, we define
\beq
\sigma_\lambda \equiv \frac{\Sigma_{\lambda\lambda}}{P_\lambda^2} ~,
\eeq
which allows us to write
\beq
\frac{b_\lambda}{a_\lambda} \sin\delta_\lambda\,\sin\phi =
-\sin^2\phi\,\sigma_\lambda ~.
\label{eq:Ssinphi}
\end{equation}

We can now express the remaining 9 observables in terms of $\Delta_i$,
$\beta + \phi$ and the newly-defined parameters $P_\lambda$,
$\xi_\lambda$ and $\sigma_\lambda$ as follows:
\bea
\label{eq:sig-perp-i}
  \Sigma_{\perp i}&=& P_i P_\perp \Bigg[\Big( \xi_\perp\,\sigma_i
   -\xi_i\,\sigma_\perp\Big)\,\cos\Delta_i-\Big(1+\xi_i\,\xi_\perp+
     \sigma_i\,\sigma_\perp\Big)\,\sin\Delta_i\Bigg]~,\\
\label{eq:lambda-perp-i}
  \Lambda_{\perp i}&=& P_i P_\perp\Bigg[\Big(\xi_\perp-\xi_i\Big)\,
     \cos\Delta_i-\Big(\sigma_i+\sigma_\perp\Big)\sin\Delta_i\,\Bigg]~,\\
\label{eq:rho-perp-i}
  \rho_{\perp i}&=& P_i P_\perp\Bigg[\Big((-1+\xi_i\,\xi_\perp+\sigma_i\,
     \sigma_\perp)\cos(2\beta+2\phi)-(\xi_i+\xi_\perp) 
     \sin(2\beta+2\phi)\Big)\cos\Delta_i\nonumber\\&&
     +\Big((-\xi_i\,\sigma_\perp+\xi_\perp\,\sigma_i )\cos(2\beta+2\phi)-
     (\sigma_i-\sigma_\perp )\sin(2\beta+2\phi)\Big)\sin\Delta_i\Bigg] ~, \\
\label{eq:sig-par-0}
  \Sigma_{\|\,0}&=& P_{\|} P_0
  \Bigg[(\xi_{\|}-\xi_{0})\sin(\Delta_0-\Delta_{\|})+(\sigma_\|
       +\sigma_0)\cos(\Delta_0-\Delta_{\|})\Bigg] ~, \\
\label{eq:lambda-par-0}
  \Lambda_{\|\,0}&=& P_{\|} P_0
  \Bigg[(\xi_{0}\sigma_{\|}-\sigma_{0}\xi_{\|})\sin(\Delta_0-
    \Delta_{\|})+(1+\xi_0\xi_{\|}+\sigma_\|\sigma_0)\cos(\Delta_0
    -\Delta_{\|})\Bigg] ~, \\
\label{eq:rho-par-0}
  \rho_{\|\,0}&=& P_{\|} P_0\Bigg[\Big((-1+\xi_{\|}\,\xi_0+\sigma_{\|}\,
     \sigma_0)\sin(2\beta+2\phi) \nn\\ 
&& \qquad\qquad
+(\xi_{\|}+\xi_0) 
     \cos(2\beta+2\phi)\Big)\cos(\Delta_0-\Delta_{\|}) \\
&& +\Big((\xi_{\|}\,\sigma_0-\xi_0\,\sigma_{\|})\sin(2\beta+2\phi)+
     (\sigma_0-\sigma_{\|})\cos(2\beta+2\phi)\Big)\sin(\Delta_0-\Delta_{\|})
  \Bigg] ~. \nn
\eea
The notable achievement of the above relations is the expression of
observables involving the interference of helicities in terms of only
3 theoretical parameters ($\Delta_i$, $\beta + \phi$), thereby
reducing the complexity of the extremization problem. The above
relations are extremely important in obtaining bounds on NP
parameters.

We now turn to the issue of new-physics signals. The presence of NP is
indicated by the violation of at least one of the relations given in
Eq.~(\ref{eq:no_np}). This in turn implies that $b_\lambda\ne 0$ and
$|\beta^{meas}_\lambda- \beta|\ne 0$ for at least one helicity
$\lambda$. Clearly, any bounds on NP parameters will depend on the
specific signal of NP. We therefore examine several different NP
signals and explore the restrictions they place on NP parameter space.

Note that we do not present an exhaustive study of new-physics
signals. The main point of the present paper is to show that it is
possible to obtain bounds on the NP parameters, even though there are
more unknowns than observables. Furthermore, the relations for the
observables are sufficiently complicated that it is not possible to
derive analytic bounds for every signal of NP. Whenever possible, we
present analytic bounds on the NP parameters. However, for certain NP
signals, we can only obtain numerical bounds. In all cases, the
bounds are found without any approximations. This demonstrates the
power of angular analysis and its usefulness in constraining NP
parameters.

We will see that, while $b_\lambda$ and $b_\lambda/a_\lambda$ can be
constrained with just one signal of NP, obtaining a bound on
$|\beta^{meas}_\lambda- \beta|$ requires at least two NP
signals. Also, because the equations are nonlinear, there are often
discrete ambiguities in the bounds. These can be reduced, leading to
stronger bounds on NP, if a larger set of observables is used.

In the subsections below we present bounds for several different
signals of NP.

\subsection{$\Sigma_{\lambda\lambda} \ne 0$}

Suppose first that one observes direct CP violation in at least one
helicity, i.e.\ $\Sigma_{\lambda\lambda} \ne 0$. The minimum value of
$b_\lambda^2$ can be obtained by minimizing $b^2_\lambda$
[Eq.~(\ref{eq:a-b-2})] with respect to $\beta$ and $\phi$:
\beq
b^2_\lambda \ge {1\over 2} \left[ \Lambda_{\lambda\lambda} -
\sqrt{\Lambda_{\lambda\lambda}^2 - \Sigma_{\lambda\lambda}^2} \right]. 
\label{eq:b-bound}
\eeq
Thus, if direct CP violation is observed, one can place a lower bound
on the new-physics amplitude $b_\lambda$.

On the other hand, it follows from Eq.~(\ref{eq:a-b-2}) that no upper
bound can ever be placed on $b_\lambda^2$. One can always take
$b_\lambda \to \infty$, as long as $\phi \to 0$ with $b_\lambda
\sin\phi$ held constant. For the same reason, one can never determine
the NP weak phase $\phi$, or place a lower bound on it. (This no
longer holds if the true value of $\beta$ is known. We discuss this
possibility in Sec.~4.6.)

It is possible, however, to place lower bounds on other NP
quantities. Using Eqs.~(\ref{eq:a-b}) and (\ref{eq:a-b-2}), it is
straightforward to obtain the constraints
\bea
\label{boundbsinphi}
&{1 \over 2} \Lambda_{\lambda\lambda} \left( 1 - y_\lambda \right) \le
b_\lambda^2\,\sin^2\phi \le {1 \over 2} \Lambda_{\lambda\lambda}
\left( 1 + y_\lambda \right) ~, & \nn\\
& {\displaystyle 1-y_\lambda \over \displaystyle1+y_\lambda} \le
r_\lambda^2 \le {\displaystyle 1+y_\lambda \over \displaystyle
1-y_\lambda} ~, & 
\eea
where
\beq
r_\lambda \equiv {b_\lambda \over a_\lambda} ~.
\label{rlambdadef}
\eeq
If $\Sigma_{\lambda\lambda} \ne 0$, these give nontrivial lower
bounds. The lower bound on $r_\lambda$ is very useful in estimating
the magnitude of NP amplitudes or the scale of NP.

One interesting observation can be made regarding bounds on
$b^2_\lambda$. Saying that new physics is present implies that the NP
amplitude $b_\lambda$ must be nonzero for at least one helicity; the
other two helicities could have vanishing NP amplitudes. A nonzero
direct asymmetry $a^{CP}_{dir}\neq 0$ (i.e.\
$\Sigma_{\lambda\lambda}\neq 0$) implies a nonzero NP amplitude with a
lower bound given by Eq.~(\ref{eq:b-bound}). Other NP signals
[Eq.~(\ref{eq:no_np})] do not bound the NP amplitude $b_\lambda^2$ for
a single helicity, but can bound combinations $(b_\lambda^2\pm
b_\sigma^2)$. This is perhaps surprising but may be understood as
follows. Consider, for example, the NP signal $\Lambda_{\perp i} \neq
0$. Even in the presence of such a signal it is possible that one of
either $b_i$ or $b_\perp$ is zero, but not both [see
Eq.~(\ref{observables})]. Thus, one can only obtain a lower bound when
simultaneously bounding $b_i^2$ and $b_\perp^2$. Hence, for
$\Lambda_{\perp i} \ne 0$, we must consider bounds on sums and
differences of the NP amplitudes, $b^2_i \pm b^2_\perp$. A similar
argument applies to all signals of NP in Eq.~(\ref{eq:no_np})
involving two helicities. We will encounter such lower bounds in
subsequent subsections.

\subsection{$\beta^{meas}_\lambda \ne \beta^{meas}_\sigma$}

Another signal of NP is if the measured value of $\beta$ is different
in two helicities, i.e.\ $\beta^{meas}_\lambda \ne
\beta^{meas}_\sigma$. We define 
\beq
2\omega_{\sigma\lambda} \equiv 2\beta^{
meas}_\sigma-2\beta^{ meas}_\lambda ~~,~~~~ \eta_\lambda \equiv
2 ( \beta^{meas}_\lambda - \beta ) ~.
\eeq
Using Eq.~(\ref{eq:a-b-2}) we have
\beq 
(b_\lambda^2\pm b_\sigma^2)\sin^2\phi =
\frac{\Lambda_{\lambda\lambda}\pm\Lambda_{\sigma\sigma}}{2}-
\frac{y_\lambda \Lambda_{\lambda\lambda}\cos\eta_\lambda\pm
y_\sigma\Lambda_{\sigma\sigma}
\cos(2\omega_{\sigma\lambda}+\eta_\lambda)}{2}~.
\label{eq:var1}
\eeq
Extremizing this expression with respect to $\eta_\lambda$, we obtain
a solution for $\eta_\lambda$:
\beq
  \label{eq:sol-eta}
\sin\eta_\lambda=\pm\frac{y_\sigma\Lambda_{\sigma\sigma}
\sin2\omega_{\sigma\lambda}}{\sqrt{y_\lambda^2
\Lambda_{\lambda\lambda}^2+y_\sigma^2\Lambda_{\sigma\sigma}^2-2\,
y_\lambda y_\sigma\Lambda_{\lambda\lambda} \Lambda_{\sigma\sigma}
\cos2\omega_{\sigma\lambda}}}
\eeq
Taking into account the sign of the second derivative, we get the
bounds
\bea
  \label{eq:bound-1}
(b_\lambda^2\pm b_\sigma^2)\sin^2\phi &\ge&
\frac{\Lambda_{\lambda\lambda}\pm\Lambda_{\sigma\sigma}}{2} -
\frac{\sqrt{y_\lambda^2 \Lambda_{\lambda\lambda}^2 +
y_\sigma^2\Lambda_{\sigma\sigma}^2\pm 2\, y_\lambda
y_\sigma\Lambda_{\lambda\lambda} \Lambda_{\sigma\sigma}
\cos2\omega_{\sigma\lambda}}}{2} ~,\\
(b_\lambda^2\pm b_\sigma^2)\sin^2\phi &\le&
\frac{\Lambda_{\lambda\lambda}\pm\Lambda_{\sigma\sigma}}{2} +
\frac{\sqrt{y_\lambda^2 \Lambda_{\lambda\lambda}^2 +
y_\sigma^2\Lambda_{\sigma\sigma}^2\pm 2\, y_\lambda
y_\sigma\Lambda_{\lambda\lambda} \Lambda_{\sigma\sigma}
\cos2\omega_{\sigma\lambda}}}{2} ~.
\label{eq:bound-2}
\eea
Extremizing with respect to $\phi$ as well, one obtains the bounds
\beq 
(b_\lambda^2 \pm b_\sigma^2) \ge \frac{\Lambda_{\lambda\lambda}
\pm \Lambda_{\sigma\sigma}}{2} - \frac{ \left\vert y_\lambda
\Lambda_{\lambda\lambda} \pm y_\sigma \Lambda_{\sigma\sigma} e^{2 i
\omega_{\sigma\lambda}} \right\vert }{2} ~,
\label{eq:bsq-omega-bounds}
\eeq
where it has been assumed that $\Lambda_{\lambda\lambda} >
\Lambda_{\sigma\sigma}$, and that the right-hand side of the inequality
is positive. (Note that an upper bound on $(b_\lambda^2 \pm
b_\sigma^2)$ cannot be obtained.) We will see below that
Eq.~(\ref{eq:bsq-omega-bounds}) plays a central role in deriving
bounds for other signals of NP.

We emphasize that all of the above bounds are exact -- no
approximations or limits have been used. {}From the constraints on
$(b_\lambda^2\pm b_\sigma^2)$ one can obtain lower bounds on
$b_\lambda^2$ and $b_\sigma^2$ individually.

Even without extremization, careful examination of Eq.~(\ref{eq:var1})
implies minimum and maximum possible values for $(b_\lambda^2\pm
b_\sigma^2)\sin^2\phi$. These can also be derived from
Eq.~(\ref{boundbsinphi}) and are given by
\bea
\label{eq:abs-min-max}
(b_\lambda^2\pm b_\sigma^2)\sin^2\phi &\ge&
\frac{\Lambda_{\lambda\lambda}\pm\Lambda_{\sigma\sigma}}{2}-
\frac{y_\lambda \Lambda_{\lambda\lambda}+
y_\sigma\Lambda_{\sigma\sigma}}{2}~,\nn\\
(b_\lambda^2\pm b_\sigma^2)\sin^2\phi &\le&
\frac{\Lambda_{\lambda\lambda}\pm\Lambda_{\sigma\sigma}}{2}+
\frac{y_\lambda \Lambda_{\lambda\lambda}+
y_\sigma\Lambda_{\sigma\sigma}}{2}~.
\eea
Note that if $2\omega_{\sigma\lambda} = 0$, Eqs.~(\ref{eq:bound-1})
and (\ref{eq:bound-2}) reproduce the bounds of
Eq.~(\ref{eq:abs-min-max}) for $(b_\lambda^2 + b_\sigma^2)\sin^2\phi$;
if $2\omega_{\sigma\lambda} = \pi$, one reproduces the bounds on
$(b_\lambda^2 - b_\sigma^2)\sin^2\phi$. If one uses other NP signals
to constrain the NP parameters, then unless these other signals result
in constraining the value of $2\omega_{\sigma\lambda}$ to be other
than 0 or $\pi$, one cannot obtain better bounds than those of
Eq.~(\ref{eq:abs-min-max}). Note also that, while
$2\omega_{\sigma\lambda}$ can be measured directly up to discrete
ambiguities, additional measurements will result in the reduction of
such ambiguities and lead to tighter bounds.

\subsection{$\Lambda_{\perp i} \ne 0$ with $\Sigma_{\lambda\lambda}=0$}

We now turn to the NP signal $\Lambda_{\perp i} \ne 0$. Here we assume
that the phase of $\bd$--$\bdbar$ mixing has not been measured in any
helicity, i.e.\ the parameter $\omega_{\perp i}$ is unknown. This
situation is plausible: as discussed above, $\Lambda_{\perp i}$ can be
obtained without tagging or time-dependence, while the measurement of
$\omega_{\perp i}$ requires both.

In order to obtain analytic bounds which depend on $\Lambda_{\perp
i}$, it is simplest to consider the limit in which all direct
CP-violating asymmetries vanish ($\Sigma_{\lambda\lambda}=0$). In this
limit, with a little algebra Eq.~(\ref{eq:lambda-perp-i}) reduces to
\beq
\frac{\Lambda_{\perp i}}{2\sqrt{\Lambda_{ii}\Lambda_{\perp\perp}}} =
-\sin\omega_{\perp i} \cos\Delta_i ~,
\label{eq:lam-perp-i}
\eeq
where $\omega_{\perp i} \equiv \beta^{ meas}_\perp -
\beta^{ meas}_i$. We solve the above for $\sin\omega_{\perp i}$
and substitute it into the expressions for $(b_i^2\pm b_\perp^2)
\sin^2\phi$ [Eq.~(\ref{eq:var1})]. The resulting expressions are
minimized straightforwardly with respect to $\cos\Delta_i$ and
$\eta_i$ to obtain new bounds. The extrema with respect to $\Delta_i$
for both $(b_i^2 \pm b_\perp^2)$ occur at
\beq
\cos^2\Delta_i = \left\{ 1 , \frac{\Lambda_{\perp
i}^2}{4\Lambda_{ii}^2\Lambda_{\perp\perp}^2\cos^2({\eta_i}/{2})} ,
\frac{\Lambda_{\perp
i}^2}{4\Lambda_{ii}^2\Lambda_{\perp\perp}^2\sin^2({\eta_i}/{2})}
\right\} ~,
\label{eq:Delext}
\eeq
while that with respect to $\eta_i$ depends on $\Lambda_{\perp i}$,
and occurs for both $(b_i^2\pm b_\perp^2)$ at
\beq
\sin\eta_i =
  \pm\frac{2R\sqrt{1-R^2}\Lambda_{\perp\perp}}{\sqrt{\Lambda_{ii}^2\pm
  2(1-2R^2)\Lambda_{ii}\Lambda_{\perp\perp}+\Lambda_{\perp\perp}^2}}
  ~,
\label{eq:etadiffext}
\eeq
where  
\beq
  \label{eq:R}
R=\frac{\Lambda_{\perp i}}{2\sqrt{\Lambda_{ii}\Lambda_{\perp\perp}}}~. 
\eeq
These extrema yield new lower limits on $(b_i^2\pm b_\perp^2)$:
\beq
2 (b_i^2 \pm b_\perp^2) \geq \Lambda_{ii} \pm \Lambda_{\perp\perp} -
\sqrt{ \left( \Lambda_{ii} \pm \Lambda_{\perp\perp}\right)^2 \mp
\Lambda_{\perp i}^2} ~,
\label{eq:bsq-Lambda-bounds}
\eeq

Interference terms such as $\Lambda_{\perp i}$ also allow us to obtain
bounds for $\eta_\lambda$. Using Eqs.~(\ref{eq:var1}) and
(\ref{eq:bsq-Lambda-bounds}), one can easily derive the bound
\bea
\label{eq:etaboundlamperpi}
(\Lambda_{ii}+ \Lambda_{\perp\perp} \cos2\omega_{\perp i})
\cos\eta_i+\Lambda_{\perp\perp}\sin2\omega_{\perp i}\sin\eta_i \leq
\sqrt{\big(\Lambda_{ii}+ \Lambda_{\perp\perp})^2-\Lambda_{\perp i}^2}
~,
\eea
which can be rewritten as
\beq
\label{eq:lam-perp-i-bound}
\Lambda_{ii} \cos\eta_i + \Lambda_{\perp\perp} \cos\eta_\perp \le
\sqrt{ \left( \Lambda_{ii} + \Lambda_{\perp\perp} \right)^2 -
\Lambda_{\perp i}^2} ~.
\eeq
Thus, if $\Lambda_{\perp i} \ne 0$, one cannot have $\eta_i =
\eta_\perp = 0$. These constraints therefore place a lower bound on
$|\beta^{meas}_i - \beta|$ and/or
$|\beta^{meas}_\perp - \beta|$.

This procedure can also be applied to $\Sigma_{\|0}$, and different
lower bounds on $(b_\|^2\pm b_0^2)$ and on $\eta_\|$, $\eta_0$ can be
derived.

Analytic bounds on $r_\lambda$ are not easy to derive, hence only
numerical bounds are obtained. We describe this in the next
subsection.

\subsection{$\Lambda_{\perp i} \ne 0$ with $\Sigma_{\lambda\lambda}\ne 0$}

We now assume that both $\Lambda_{\perp i} \ne 0$ and
$\Sigma_{\lambda\lambda}\ne 0$, but no measurement has been made of
the parameter $\omega_{\perp i}$. In this case the procedure outlined
in the previous subsection cannot be used to obtain analytic bounds on
$(b_i^2 \pm b_\perp^2)$. The reason is that one does not find a simple
solution for $\omega_{\perp i}$ such as that given in
Eq.~(\ref{eq:lam-perp-i}).  In this case, we are forced to turn to
numerical solutions. We use the same method as in the previous
subsection --- we solve Eq.~(\ref{eq:lambda-perp-i}) for
$\omega_{\perp i}$ and substitute it into Eq.~(\ref{eq:var1}) ---
except that now the minimization is performed numerically with respect
to the variables $\eta_i$, $\phi$ and $\Delta_i$ using the computer
program MINUIT \cite{minuit}.

\begin{figure}
\centerline{\epsfxsize 3.5 truein \epsfbox {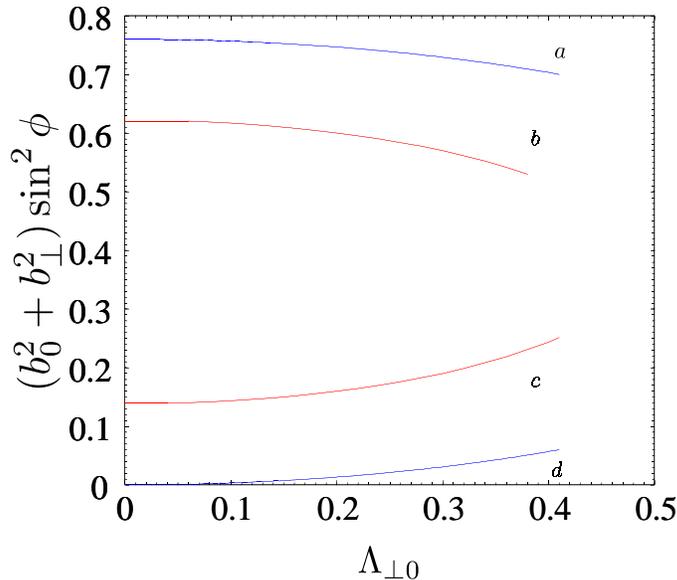}}
\caption{The lower and upper bounds on $(b^2_0+b^2_\perp)\sin^2\phi$
as a function of $\Lambda_{\perp 0}$. For curves $b$ and $c$ we have
assumed the following values for the observables: $\Lambda_{00}=0.6$,
$\Lambda_{\perp\perp}=0.16$, $y_0=0.60$, $y_\perp=0.74$. Curves $a$
and $d$ represent the corresponding case with no direct CP asymmetry
(i.e.\ $y_0 = y_\perp = 1.0$).}
\label{Fig1}
\end{figure}

We assume the new-physics signal $\Lambda_{\perp 0} \ne 0$. In order
to perform numerical minimization, we must choose values for the
observables. Here and in the next subsection, we take
$\Lambda_{00}=0.6$, $\Lambda_{\perp \perp}=0.16$, $y_0=0.60$ and
$y_\perp=0.74$.

In Fig.~\ref{Fig1}, we present the lower and upper bounds on $(b_0^2 +
b_\perp^2) \sin^2\phi$ as a function of $\Lambda_{\perp 0}$. As in the
previous subsection, these bounds are obtained by minimizing with
respect to the variables $\Delta_i$ and $\eta_i$. Since the minimum
value of $(b^2_0+b^2_\perp)$ can be obtained from that of $(b_0^2 +
b_\perp^2) \sin^2\phi$ by setting $\sin\phi = 1$ (its maximum value),
the lower bound on $(b^2_0+b^2_\perp)$ is identical to that of $(b_0^2
+ b_\perp^2) \sin^2\phi$. However, upper bounds can only be derived
for $(b^2_0+b^2_\perp)\sin^2\phi$. For comparison, we include the
bounds for the case of vanishing direct CP asymmetry, i.e.\
$\Sigma_{00} = \Sigma_{\perp\perp} = 0$
[Eq.~(\ref{eq:bsq-Lambda-bounds})]. It is clear that the bounds are
stronger if there are more signals of NP.

\begin{figure}
\centerline{\epsfxsize 3.5 truein \epsfbox {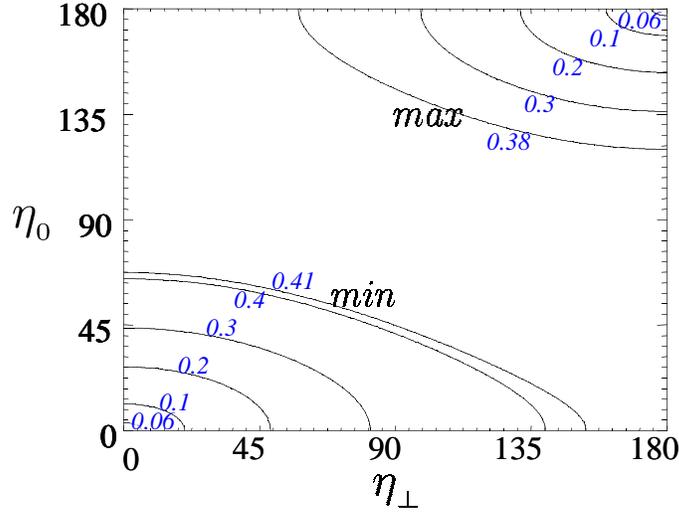}}
\caption{Contours showing the (correlated) lower and upper bounds on
$\eta_0$ and $\eta_\perp$, corresponding to the different values of
$\Lambda_{\perp 0}$ shown on the Figure. We have assumed the following
values for the observables: $\Lambda_{00}=0.6$,
$\Lambda_{\perp\perp}=0.16$, $y_0=0.60$, $y_\perp=0.74$. Values of
$\eta_0$ and $\eta_\perp$ above (below) and to the right (left) of the
minimum (maximum) contours are allowed.}
\label{Fig2}
\end{figure}

As in the previous subsection, the constraints on $(b_0^2 +
b_\perp^2)\sin^2\phi$ imply certain allowed regions for $\eta_0$ and
$\eta_\perp$ (see Eq.~(\ref{eq:lam-perp-i-bound}) and the surrounding
discussion). These are shown in Fig.~\ref{Fig2}. Recall that
$\eta_\lambda \equiv 2 ( \beta^{meas}_\lambda - \beta )$.
Since it is not possible to simultaneously have $\eta_0 = \eta_\perp =
0$ (or $\pi$), this is a clear sign of NP (as is $\Lambda_{\perp 0}
\ne 0$). However, since neither $\eta_0$ nor $\eta_\perp$ is
constrained to lie within a certain range, no bounds on $\beta$ can be
derived.

One can perform a similar numerical extremization for $(b_0^2 -
b_\perp^2)\sin^2\phi$. However, for this particular data set, we
simply reproduce the bounds of Eq.~(\ref{eq:abs-min-max}): $-0.02 \le
(b_0^2 - b_\perp^2)\sin^2\phi \le 0.46$. Since this bound is
independent of $\Lambda_{\perp 0}$, we have not plotted it.

The easiest way to see whether the numerical extremization of $(b_0^2
\pm b_\perp^2)\sin^2\phi$ depends on $\Lambda_{\perp 0}$ or not is as
follows. We refer to Eq.~(\ref{eq:var1}), and note that
$2\omega_{\perp 0} + \eta_0 = \eta_\perp$. The minimal [maximal] value
of $(b_0^2 + b_\perp^2)\sin^2\phi$ occurs at the point $(\eta_0,
\eta_\perp) = (0,0)$ [$(\pi,\pi)$]. Thus, the minimal [maximal] value
of $(b_0^2 + b_\perp^2)\sin^2\phi$ depends on $\Lambda_{\perp 0}$ only
if the point $(0,0)$ [$(\pi,\pi)$] is excluded. Similarly, the minimal
[maximal] value of $(b_0^2 - b_\perp^2)\sin^2\phi$ depends on
$\Lambda_{\perp 0}$ only if the point $(0,\pi)$ [$(\pi,0)$] is
excluded. Referring to Fig.~\ref{Fig2}, we note that the points
$(\eta_0, \eta_\perp) = (0,0)$, $(\pi, \pi)$ are excluded. Thus, the
minimal and maximal values of $(b_0^2 + b_\perp^2)\sin^2\phi$ depend
on $\Lambda_{\perp 0}$, as in Fig.~\ref{Fig1}. On the other hand, the
points $(\eta_0, \eta_\perp) = (0,\pi)$ and $(\pi,0)$ are allowed, so
the minimal and maximal values of $(b_0^2 - b_\perp^2)\sin^2\phi$ are
independent of $\Lambda_{\perp 0}$, as described above.

As noted previously, the minimal values for $(b_0^2 \pm b_\perp^2)$
are equal to those for $(b_0^2 \pm b_\perp^2) \sin^2\phi$. These
values can then be combined to give individual minima on $b_0^2$ and
$b_\perp^2$.

\begin{figure}
\centerline{\epsfxsize 3.0 truein \epsfbox {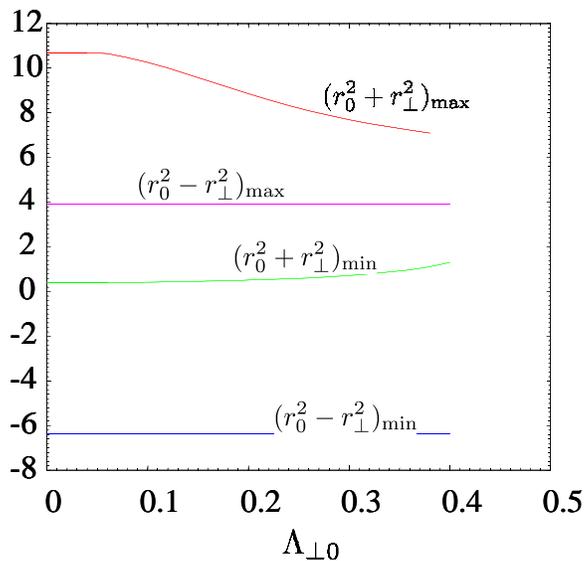}}
\caption{Upper and lower bounds on $r^2_0\pm r^2_\perp$ as a function
  of $\Lambda_{\perp 0}$. We have assumed the following values for the
  observables: $\Lambda_{00}=0.6$, $\Lambda_{\perp \perp}=0.16$,
  $y_0=0.60$, $y_\perp=0.74$.}
\label{Fig3}
\end{figure}

It is also possible to obtain numerical bounds on the combinations of
ratios $r^2_0\pm r^2_\perp$ [Eq.~(\ref{rlambdadef})]. The procedure is
very similar to that used to obtain bounds on $(b_0^2 \pm b_\perp^2)
\sin^2\phi$. The bounds on $r^2_0\pm r^2_\perp$ are shown in
Fig.~\ref{Fig3}. As was the case for $(b_0^2 - b_\perp^2)\sin^2\phi$,
the bounds on $r^2_0-r^2_\perp$ are independent of $\Lambda_{\perp i}$
and follow directly from Eq.~(\ref{boundbsinphi}): $-6.44 \le
r^2_0-r^2_\perp \le 3.85$. However, unlike $b^2_0\pm b^2_\perp$, upper
bounds on $r^2_0\pm r^2_\perp$ can also be obtained. The upper and
lower bounds on $r_0^2\pm r_\perp^2$ can then be used to bound $r_0^2$
and $r_\perp^2$ individually. This constrains the scale of new
physics, and so is very significant.

\subsection{Observation of $\Lambda_{\perp 0}$ and $\Sigma_{\perp 0}$ with
$\Sigma_{00}\ne 0$, $\Sigma_{\perp\perp}\ne 0$.}

In this subsection we assume that, in addition to $\Lambda_{\perp 0}$,
$\Sigma_{\perp 0}$ is also known ($\omega_{\perp 0}$ is still assumed
not to have been measured). We then see, from
Eqs.~(\ref{eq:sig-perp-i}) and (\ref{eq:lambda-perp-i}), that both
$\cos(\Delta_0)$ and $\sin(\Delta_0)$ can be determined in terms of
these two observables.  Thus, $\Delta_0$ can be obtained without
ambiguity. Furthermore, using the relation
$\cos^2(\Delta_0)+\sin^2(\Delta_0)=1$, we can {\it solve} for
$\omega_{\perp 0}$, up to an 8-fold discrete ambiguity (i.e.\ a 4-fold
ambiguity in $2\omega_{\perp 0}$)\footnote{It is to be expected that
we can solve for $\omega_{\perp 0}$ in this case.  If the theoretical
parameters $\beta$ and $\phi$ did not vanish from the equation
$\cos^2(\Delta_0)+\sin^2(\Delta_0)=1$, then we would have a relation
between the {\it independent} parameters $\beta$ and $\phi$, which is
impossible. $\beta$ and $\phi$ are eliminated because this equation
depends on $(2\beta^{meas}_\perp-2\beta-2\phi) -
(2\beta^{meas}_0-2\beta-2\phi) = 2\omega_{\perp 0}$.}. This is shown
explicitly in Appendix 1. Thus, $\omega_{\perp 0}$ does not take a
range of values, as in the previous subsections, but instead takes
specific values. (In fact, one can solve for $\omega_{\perp 0}$, up to
discrete ambiguities, whenever two observables are known which involve
the interference of two helicity amplitudes.)

The expressions and values for $\Delta_0$ and $\omega_{\perp 0}$ are
then substituted into Eq.~(\ref{eq:var1}), and we use MINUIT to
numerically minimize the resulting expression with respect to $\eta_i$
and $\phi$.  As before, we take $\Lambda_{00}=0.6$, $\Lambda_{\perp
\perp}=0.16$, $y_0=0.60$ and $y_\perp=0.74$.

\begin{figure}
\vskip -1.0truein
\centerline{\epsfxsize 4.5 truein \epsfbox {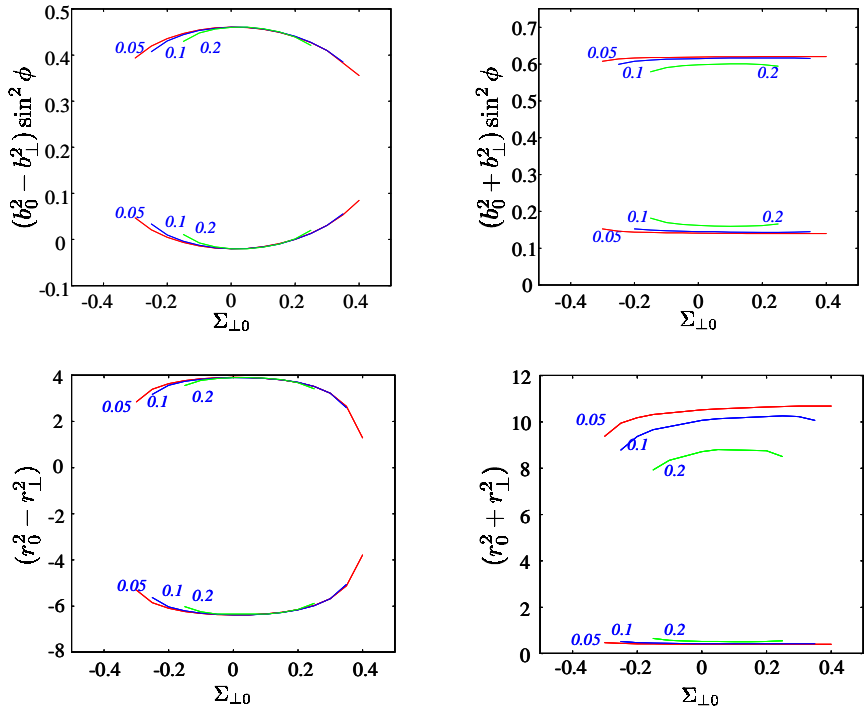}}
\caption{The lower and upper bounds on $(b^2_0 \pm b^2_\perp)
\sin^2\phi$ and $(r^2_0 \pm r^2_\perp)$ as a function of
$\Sigma_{\perp 0}$. Each curve corresponds to a specific value of
$\Lambda_{\perp 0}$, shown on the Figure. We have assumed the
following values for the observables: $\Lambda_{00}=0.6$,
$\Lambda_{\perp\perp}=0.16$, $y_0=0.60$, $y_\perp=0.74$.}
\label{Fig4}
\end{figure}

The numerical constraints on $(b^2_0 \pm b^2_\perp)\sin^2\phi$ and
$(r^2_0 \pm r^2_\perp)$ are shown in Fig.~\ref{Fig4}. In these
figures, we have only presented results for positive values of
$\Lambda_{\perp 0}$. A point on a plot with a negative value of
$\Lambda_{\perp 0}$ is equivalent to that with a positive
$\Lambda_{\perp 0}$ and negative $\Sigma_{\perp 0}$. This interchange
reverses the signs of $\cos(\Delta_0)$ and $\sin(\Delta_0)$, but does
not change the value of $\omega_{\perp 0}$.

\begin{figure}
\centerline{\epsfxsize 4.5 truein \epsfbox {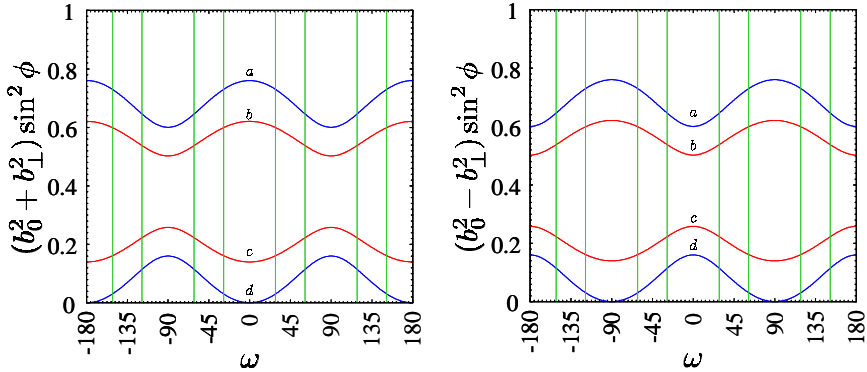}}
\caption{The lower and upper bounds on $(b^2_0 \pm b^2_\perp)
\sin^2\phi$ as a function of $\omega_{\perp 0}$. For curves $b$ and
$c$ we have assumed the following values for the observables:
$\Lambda_{00}=0.6$, $\Lambda_{\perp\perp}=0.16$, $y_0=0.60$,
$y_\perp=0.74$. Curves $a$ and $d$ represent the corresponding case
with no direct CP asymmetry (i.e.\ $y_0 = y_\perp = 1.0$). The
solutions for $\omega_{\perp 0}$ for $\Lambda_{\perp 0}=0.2$ and
$\Sigma_{\perp 0}=0.2$ are shown as vertical lines.}
\label{Fig5}
\end{figure}

As noted above, the knowledge of both $\Lambda_{\perp 0}$ and
$\Sigma_{\perp 0}$ allows us to fix the value of $\omega_{\perp 0}$,
up to an 8-fold discrete ambiguity. In this case, we can use
Eqs.~(\ref{eq:bound-1}), (\ref{eq:bound-2}) and
(\ref{eq:bsq-omega-bounds}) to directly bound $(b_\lambda^2\pm
b_\sigma^2)\sin^2\phi$. This is illustrated in Fig.~\ref{Fig5} for
$\Lambda_{\perp 0} = 0.2$ and $\Sigma_{\perp 0} = 0.2$.

Of course, it is also possible to measure $2\omega_{\perp 0}$ directly
[Eq.~(\ref{betameasdef})], up to a 4-fold discrete ambiguity. As we
show in Appendix 1, in general these four values only partially
overlap with the four values obtained from the derivation of
$2\omega_{\perp 0}$ from measurements of $\Lambda_{\perp 0}$ and
$\Sigma_{\perp 0}$ -- the discrete ambiguity in $2\omega_{\perp 0}$ is
reduced to twofold. Thus, by combining the two ways of obtaining
$2\omega_{\perp 0}$, the discrete ambiguity can be reduced. This will
in turn improve the bounds on the NP parameters.

As in the previous subsection, one can also place (correlated)
constraints on $\eta_0$ and $\eta_\perp$. In itself, this does not
lead to a bound on $\beta$. However, if in addition
$2\,\beta^{meas}_{\lambda}$ is measured directly
[Eq.~(\ref{betameasdef})], then $\beta$ can be constrained.

\subsection{Measurement of $\beta$}

Finally, suppose that an angular analysis of $\bd(t) \to J/\psi
K^{*0}$ is done, and no new physics is found. This implies that the
true $\bd$--$\bdbar$ mixing phase $\beta$ can be extracted from
measurements of CP violation in this decay. Now suppose that some NP
signal is found in $\bd(t) \to \phi K^{*0}$. The analysis described in
the previous sections can now be applied, except that in this case we
{\it know} the value of $\beta$. In addition to improving bounds on
$b^2_\lambda$ and $r^2_\lambda$ using previous techniques, we can now
constrain the NP phase $\phi$.

For example, assuming that $\beta$ is known, one can use
Eq.~(\ref{eq:a-b-2}) to improve the bound on $b_\lambda^2$:
\beq
b_\lambda^2 \ge \frac{\Lambda_{\lambda\lambda}}{2}
\Big(1-y_\lambda\cos(2\beta^{meas}_\lambda-2\beta)\Big) ~.
\label{eq:b-beta-bound}
\eeq
$2\beta^{meas}_\lambda$ and $2\beta$ can each be obtained with a
twofold ambiguity. Their combination leads to a twofold ambiguity in
the bound for $b^2_\lambda$. Obviously, to be conservative, we take
the weaker of the two bounds.

To obtain a meaningful bound on $\phi$, we require the use of
$r^2_\lambda$. In previous subsections we have derived bounds on
$(r^2_0\pm r^2_\perp)$ (Figs.~\ref{Fig3} and \ref{Fig4}). Bounds on
$(r^2_\|\pm r^2_\perp)$ and $(r^2_0\pm r^2_\|)$ can also be obtained.
These can all be combined to yield upper and lower bounds on
$r_\lambda^2$. Together, Eqs.~(\ref{eq:a-b}) and (\ref{eq:a-b-2})
provide a constraint on $\phi$:
\beq
(r_\lambda^2)_{\rm min}\leq \frac{1-y_0\cos(2\beta^{meas}_0-2\beta)}
{1-y_0\cos(2\beta^{meas}_0-2\beta-2\phi)}\leq (r_\lambda^2)_{\rm max}
~.
\label{eq:r-phi-bound}
\eeq
In this case, there is an eightfold ambiguity on the bounds on $\sin
2\phi$.

\section{Discussion \& Summary}

In this paper we consider $\bvv$ decays in which ${\overline{V}}_1
{\overline{V}}_2 = V_1 V_2$, so that both $B^0$ and ${\bar B}^0$ can
decay to the final state $V_1 V_2$. If a time-dependent angular
analysis of $B^0(t) \to V_1 V_2$ can be performed, it is possible to
extract 18 observables [Eq.~(\ref{eq:obs})]. However, there are only
six helicity amplitudes describing the decays $\bvv$ and ${\bar B} \to
V_1 V_2$. There are therefore only 11 independent observables
(equivalent to the magnitudes and relative phases of the six helicity
amplitudes).

We assume that the $\bvv$ decays are dominated by a single decay
amplitude in the standard model (SM). The SM parametrization of such
decays contains six theoretical parameters: three helicity amplitudes
$a_\lambda$, two relative strong phases, and the weak phase $\beta$
(the phase of $\bd$--$\bdbar$ mixing). Because there are 18
observables, one has a total of 12 relations to test for the presence
of new physics (NP) [Eq.~(\ref{eq:no_np})]. With 11 independent
observables and six SM parameters, one might expect that only five
tests are necessary to search for NP. However, because the equations
relating the observables to the theoretical parameters are nonlinear
[Eq.~(\ref{observables})], for certain (fine-tuned) values of the SM
parameters, some tests can agree with the SM predictions, even in the
presence of NP. To take this possibility into account, all 12 NP tests
are needed to perform a complete search for NP.

In this paper we assume that a single NP amplitude contributes to
$\bvv$ decays. In this case one finds a total of 13 theoretical
parameters: in addition to the six SM parameters, there are three NP
helicity amplitudes $b_\lambda$, three additional relative strong
phases, and one NP weak phase $\phi$. Suppose now that a NP signal is
seen. With only 11 independent observables, it is clear that one
cannot extract any of the NP parameters. However, precisely because
the equations in Eq.~(\ref{observables}) are nonlinear, one can place
{\it lower bounds} on the theoretical parameters. This is the main
point of the paper.

In the previous section we presented several such constraints, which
we summarize here. The form of the constraints depends on which
observables have been measured. In some cases, it is possible to
obtain analytic results; in other cases only numerical bounds are
possible. 

For example, three distinct NP signals are $\Sigma_{\lambda\lambda}
\ne 0$, $\beta^{meas}_\lambda \ne
\beta^{meas}_\sigma$, and $\Lambda_{\perp i} \ne 0$ (with
$\Sigma_{\lambda\lambda}=0$). In all three cases one can derive
analytic lower bounds on the size of $b_\lambda$:
\bea
b^2_\lambda & \ge & {1\over 2} \left[ \Lambda_{\lambda\lambda} -
\sqrt{\Lambda_{\lambda\lambda}^2 - \Sigma_{\lambda\lambda}^2} \right], \nn\\
(b_\lambda^2 \pm b_\sigma^2) & \ge & \frac{\Lambda_{\lambda\lambda}
\pm \Lambda_{\sigma\sigma}}{2} - \frac{ \left\vert y_\lambda
\Lambda_{\lambda\lambda} \pm y_\sigma \Lambda_{\sigma\sigma} e^{2 i
\omega_{\sigma\lambda}} \right\vert }{2} ~, \nn\\
2 (b_i^2 \pm b_\perp^2) & \ge & \Lambda_{ii} \pm \Lambda_{\perp\perp}
- \sqrt{ \left( \Lambda_{ii} \pm \Lambda_{\perp\perp}\right)^2 \mp
\Lambda_{\perp i}^2} ~,
\eea
where $y_\lambda \equiv \sqrt{1 -
\Sigma_{\lambda\lambda}^2/\Lambda_{\lambda\lambda}^2}$ and
$2\omega_{\sigma\lambda} \equiv 2\beta^{
meas}_\sigma-2\beta^{ meas}_\lambda$.  A-priori, one does not
know which of the above constraints will be strongest -- this will
depend on the measured values of the observables and/or which NP
signals are seen.

Constraints on other theoretical parameters are possible. For example,
if one measures $\Lambda_{\perp i} \ne 0$ (with
$\Sigma_{\lambda\lambda}=0$), one finds
\beq
\Lambda_{ii} \cos\eta_i + \Lambda_{\perp\perp} \cos\eta_\perp \le
\sqrt{ \left( \Lambda_{ii} + \Lambda_{\perp\perp} \right)^2 -
\Lambda_{\perp i}^2} ~,
\eeq
where $\eta_\lambda \equiv 2 ( \beta^{meas}_\lambda - \beta)$. Thus,
if $\Lambda_{\perp i} \ne 0$, one obtains correlated lower bounds on
$|\beta^{meas}_i - \beta|$ and $|\beta^{meas}_\perp - \beta|$.

If more observables or NP signals are measured, then it is not
possible to obtain analytic constraints -- one must perform a
numerical analysis. In Sec.~4.4 we presented numerical results for
$\Lambda_{\perp 0} \ne 0$ with $\Sigma_{00}\ne 0$ and
$\Sigma_{\perp\perp}\ne 0$. In Sec.~4.5 we assumed that in addition
$\Sigma_{\perp 0}$ was measured. In both cases we were able to put
lower bounds on $(b_0^2 \pm b_\perp^2)$. (Upper bounds are possible
only for $(b_0^2 + b_\perp^2) \sin^2\phi$.) We also obtained bounds on
$r^2_0\pm r^2_\perp$ ($r_\lambda \equiv b_\lambda / a_\lambda$).

The bounds improve as more NP signals are included in the fits. This
is logical. For a particular NP signal, the bounds are weakest if that
signal is zero. (Indeed, the bounds vanish if all NP signals are
zero.) If a nonzero value for that signal is found, the bound will
improve. Similarly, the bounds generally improve if additional
observables are measured, even if they are not signals of NP. This is
simply because additional measurements imply additional constraints,
which can only tighten bounds on the theoretical parameters.


This behaviour is seen most clearly in Secs.~4.3--4.5. Consider the
lower bound on $(b_0^2 + b_\perp^2) \sin^2\phi$ as a function of
$\Lambda_{\perp 0}$. In Sec.~4.3, it is assumed that the NP signal
$\Sigma_{\lambda\lambda} = 0$. In Fig.~\ref{Fig1} we see that the
bound is strengthened, varying from 0 ($\Lambda_{\perp 0} = 0$) to
about 0.05 ($\Lambda_{\perp 0} = 0.4$). In Sec.~4.4, the values
$y_0=0.60$ and $y_\perp=0.74$ are taken, i.e.\ it is assumed that both
NP signals $\Sigma_{00}$ and $\Sigma_{\perp\perp}$ are nonzero. In
this case Fig.~\ref{Fig1} shows that the lower bound varies from 0.14
($\Lambda_{\perp 0} = 0$) to 0.24 ($\Lambda_{\perp 0} = 0.4$). For
$\Lambda_{\perp 0} = 0.2$, the bound is 0.16. In Sec.~4.5 the
measurement of $\Sigma_{\perp 0}$ (not a NP signal) is added. Now the
lower bound on $(b_0^2 + b_\perp^2) \sin^2\phi$ depends on the values
of both $\Lambda_{\perp 0}$ and $\Sigma_{\perp 0}$. {}From
Fig.~\ref{Fig4}, we see that it takes the value 0.18 for
$\Lambda_{\perp 0} = 0.2$ and $\Sigma_{\perp 0} = -0.15$.

In addition to the bounds on the $b_\lambda$ and $r_\lambda$, it is
possible to find correlated numerical constraints on the
$\eta_\lambda$, as in Fig.~\ref{Fig2}. If these are combined with a
measurement of $2\,\beta^{meas}_{\lambda}$, one can then obtain a
bound on $\beta$, even though NP is present.

Even if $2\omega_{\sigma\lambda}$ is not measured directly, one can
obtain its value (up to a fourfold ambiguity) through measurements of
two observables involving the interference of two helicity amplitudes
(as well as the $\Lambda_{\lambda\lambda}$ and
$\Sigma_{\lambda\lambda}$). These can be converted into bounds on the
other NP parameters. If $2\omega_{\sigma\lambda}$ is measured
directly, this reduces the discrete ambiguity to twofold, and improves
the bounds.

Finally, all of the above bounds assume that the true (SM) value of
$\beta$ is not known. However, it is possible that no NP is seen in
$\bd(t) \to J/\psi K^{*0}$, in which case measurements of CP violation
in this decay allow one to extract the true value of $\beta$. This
value of $\beta$ can then be used as an {\it input} to the analysis of
other decays, such as $\bd(t) \to \phi K^{*0}$, in which NP signals
might be found. If $\beta$ is assumed to be known, then the bounds on
$b_\lambda^2$ and $r_\lambda^2$ described above are tightened, in
general. In addition, it is possible to place bounds on the NP weak
phase $\phi$.

We stress that we have not presented a complete list of constraints on
the NP parameters -- the main aim of this paper was simply to show
that such bounds exist. Our results have assumed that only a subset of
all observables has been measured, and the bounds vary depending on
the NP signal found. In practice, the constraints will be obtained by
performing a numerical fit using all measurements. If it is possible
to measure all observables, one will obtain the strongest constraints
possible.

As a specific application, we have noted the apparent discrepancy in
the value of $\sin 2\beta$ as obtained from measurements of $\bd(t)\to
J/\psi \ks$ and $\bd(t) \to \phi \ks$. In this case, the angular
analyses of $\bd(t)\to J/\psi K^*$ and $\bd(t) \to \phi K^*$ would
allow one to determine if new physics is indeed present. If NP is
confirmed, the method described in this paper would allow one to put
constraints on the NP parameters. If NP is subsequently discovered in
direct searches at the LHC or GLC, these bounds would indicate whether
this NP could be responsible for that seen in $B$ decays.

\bigskip
\noindent
{\bf Acknowledgements}:
N.S. and R.S. thank D.L. for the hospitality of the Universit\'e de
Montr\'eal, where part of this work was done. The work of D.L. was
financially supported by NSERC of Canada. The work of Nita Sinha was
supported by a project of the Department of Science and Technology,
India, under the young scientist scheme.

\section*{Appendix 1}

Assume that, in addition to $\Lambda_{00}$, $\Lambda_{\perp\perp}$,
$\Sigma_{00}$ and $\Sigma_{\perp\perp}$, $\Lambda_{\perp 0}$ and
$\Sigma_{\perp 0}$ are also known. The expressions for these last two
quantities are (repeated for convenience)
\bea
\label{sigmadef}
\Sigma_{\perp 0}&=& P_0 P_\perp \Bigg[\Big( \xi_\perp\,\sigma_0
  -\xi_0\,\sigma_\perp\Big)\,\cos\Delta_0-\Big(1+\xi_0\,\xi_\perp+
  \sigma_0\,\sigma_\perp\Big)\,\sin\Delta_0\Bigg]~,\\
\label{lambdadef}
\Lambda_{\perp 0}&=& P_0 P_\perp\Bigg[\Big(\xi_\perp-\xi_0\Big)\,
\cos\Delta_0-\Big(\sigma_0+\sigma_\perp\Big)\sin\Delta_0\,\Bigg]~,
\eea
where
\bea
\label{defs}
P_\lambda^2 &\equiv& \Lambda_{\lambda\lambda}[1-y_\lambda\,
       \cos (2\beta^{meas}_\lambda-2\beta-2\phi)] ~, \nn\\
\xi_\lambda &\equiv& \frac{\Lambda_{\lambda\lambda}\, y_\lambda\,
       \sin (2\beta^{meas}_\lambda-2\beta-2\phi)}{P_\lambda^2} ~, \nn\\
\sigma_\lambda & \equiv & \frac{\Sigma_{\lambda\lambda}}{P_\lambda^2}
~,
\eea
with
\beq
\label{ydef}
y_\lambda \equiv \sqrt{1 - {\Sigma_{\lambda\lambda}^2 \over
\Lambda_{\lambda\lambda}^2}} ~.
\eeq

Eqs.~(\ref{sigmadef}) and (\ref{lambdadef}) can be solved for
$\cos\Delta_i$ and $\sin\Delta_i$. Writing
\bea
\Lambda_{\perp 0} & = & A \cos\Delta_0 + B \sin\Delta_0 ~, \nn\\
\Sigma_{\perp 0} & = & A' \cos\Delta_0 + B' \sin\Delta_0 ~,
\eea
where
\bea
A \equiv P_0 P_\perp (\xi_\perp - \xi_0) & , & ~~ B \equiv - P_0
P_\perp (\sigma_0 + \sigma_\perp) ~, \nn\\
A' \equiv P_0 P_\perp (\xi_\perp \sigma_0 - \xi_0 \sigma_\perp)
& , &~~ B' \equiv - P_0 P_\perp (1 + \xi_0 \xi_\perp + \sigma_0
\sigma_\perp ) ~,
\eea
we get
\beq
\cos\Delta_0 = {B' \Lambda_{\perp 0} - B \Sigma_{\perp 0} \over A B' -
B A'} ~,~~
\sin\Delta_0 = {A' \Lambda_{\perp 0} - A \Sigma_{\perp 0} \over A' B -
B' A} ~.
\eeq
Then the relation $\cos^2 \Delta_0 + \sin^2 \Delta_0 = 1$ results in
\beq
1 = { ({A'}^2 + {B'}^2) \, \Lambda_{\perp 0}^2 + ({A}^2 +
{B}^2) \, \Sigma_{\perp 0}^2 - 2 (A A' + B B') \, \Lambda_{\perp 0}
\, \Sigma_{\perp 0} \over (A B' - B A')^2} ~.
\eeq
The point is that each of the four combinations $({A}^2 + {B}^2)$,
$({A'}^2 + {B'}^2)$, $(A A' + B B')$ and $(A B' - B A')$, is
independent of $\beta$ and $\phi$.

In order to show this, the following relations are useful:
\bea
\xi_\lambda^2 & = & -\sigma_\lambda^2 + 2 {\Lambda_{\lambda\lambda}
\over P_\lambda^2} - 1 ~, \nn\\
\xi_0 \xi_\perp & = & {\Lambda_{00} \Lambda_{\perp\perp} y_0 y_\perp
\over P_0^2 P_\perp^2} \cos 2 \omega_{\perp 0} - 1 -
{\Lambda_{00} \Lambda_{\perp\perp} \over P_0^2 P_\perp^2} +
{\Lambda_{00} \over P_0^2} + {\Lambda_{\perp\perp} \over P_\perp^2} ~,
\nn\\
P_\lambda^4 & = & - \Lambda_{\lambda\lambda}^2 + 2
\Lambda_{\lambda\lambda} P_\lambda^2 + \Lambda_{\lambda\lambda}^2
y_\lambda^2 \cos^2 (2\beta^{meas}_\lambda-2\beta-2\phi) ~,
\eea
where
\beq
2\omega_{\perp 0} \equiv 2\beta^{ meas}_\perp-2\beta^{
meas}_0 ~.
\eeq
With these one can show that
\bea
(A^2 + B^2) & = & 2 \Lambda_{00} \Lambda_{\perp\perp} + 2 \Sigma_{00}
\Sigma_{\perp\perp} - 2 \Lambda_{00} \Lambda_{\perp\perp} y_0 y_\perp
\cos 2\omega_{\perp 0} ~, \nn\\
({A'}^2 + {B'}^2) & = & 2 \Lambda_{00} \Lambda_{\perp\perp} + 2
\Sigma_{00} \Sigma_{\perp\perp} + 2 \Lambda_{00} \Lambda_{\perp\perp}
y_0 y_\perp \cos 2\omega_{\perp 0} ~, \nn\\
(A A' + B B') & = & 2 ( \Lambda_{\perp\perp} \Sigma_{00} + \Lambda_{00}
\Sigma_{\perp\perp} ) ~, \nn\\
(A B' - B A')^2 & = & 4 \left( \Lambda_{00}^2 - \Sigma_{00}^2 \right)
\left( \Lambda_{\perp\perp}^2 - \Sigma_{\perp\perp}^2 \right) \sin^2
2\omega_{\perp 0} ~.
\eea

Thus, the relation $\cos^2 \Delta_0 + \sin^2 \Delta_0 = 1$ gives a
quadratic equation in $\cos 2\omega_{\perp 0}$:
\beq
N_1 (1 - \cos^2 2\omega_{\perp 0}) = (N_2 + M_2 \cos 2 \omega_{\perp
0}) + (N_3 + M_3 \cos 2 \omega_{\perp 0}) + N_4 ~,
\eeq
where
\bea
N_1 = 4 \left( \Lambda_{00}^2 - \Sigma_{00}^2 \right) \left(
\Lambda_{\perp\perp}^2 - \Sigma_{\perp\perp}^2 \right) & , & \nn\\
N_2 = 2 \left( \Lambda_{00} \Lambda_{\perp\perp} + 2 \Sigma_{00}
\Sigma_{\perp\perp} \right) \Lambda_{\perp 0}^2 & , & M_2 = 2
\Lambda_{00} \Lambda_{\perp\perp} y_0 y_\perp \Lambda_{\perp 0}^2 ~,
\nn\\
N_3 = 2 \left( \Lambda_{00} \Lambda_{\perp\perp} + 2 \Sigma_{00}
\Sigma_{\perp\perp} \right) \Sigma_{\perp 0}^2 & , & M_3 = - 2
\Lambda_{00} \Lambda_{\perp\perp} y_0 y_\perp \Sigma_{\perp 0}^2 ~,
\nn\\
N_4 = - 2 \left( \Lambda_{\perp\perp} \Sigma_{00} + \Lambda_{00}
\Sigma_{\perp\perp} \right) \Lambda_{\perp 0} \Sigma_{\perp 0} & . &
\eea
The solution for $\cos 2\omega_{\perp 0}$ is
\beq
\cos 2\omega_{\perp 0} = { - (M_2 + M_2) \pm \sqrt{ (M_2 + M_3)^2 - 4
N_1 (N_2 + N_3 + N_4 - N_1) } \over 2 N_1} ~.
\eeq
Thus, we obtain $2\omega_{\perp 0}$ with a 4-fold discrete ambiguity
(or, equivalently, $\omega_{\perp 0}$ with an 8-fold ambiguity).

It is also possible to obtain $2\omega_{\perp 0}$ from direct
measurements of $\rho_{00}$ and $\rho_{\perp\perp}$
[Eq.~(\ref{betameasdef})]. However, it is $\sin\,
2\,\beta^{meas}_{\lambda}$ which is measured, so that one extracts two
values:
\beq
2\,\beta^{meas}_{\lambda} ~~,~~~~ \pi - 2\,\beta^{meas}_{\lambda} ~.
\eeq
This leads to a 4-fold discrete ambiguity in $2\omega_{\perp 0}$:
\beq
2\omega_{\perp 0} = \pm (2\beta^{ meas}_\perp-2\beta^{
meas}_0) ~~,~~~~ \pm (2\beta^{ meas}_\perp + 2\beta^{
meas}_0 - \pi) ~.
\eeq

Of these four values, in general only two will be found among those
obtained by deriving $2\omega_{\perp 0}$ from measurements of
$\Lambda_{00}$, $\Lambda_{\perp\perp}$, $\Sigma_{00}$,
$\Sigma_{\perp\perp}$, $\Lambda_{\perp 0}$, and $\Sigma_{\perp 0}$.
Thus, by extracting $2\omega_{\perp 0}$ in these two different ways,
one can reduce the discrete ambiguity to twofold.

Note that this can only be done if new physics is found. If no NP
signal is observed, then $2\omega_{\perp 0} = 0$, and discrete
ambiguities are irrelevant.


\end{document}